\documentclass[twocolumn,showpacs,superscriptaddress]{revtex4}
\usepackage{epsfig}
\usepackage{graphicx}
\usepackage{epstopdf}
\usepackage{amssymb}
\usepackage{color}
\usepackage{subfigure}
\usepackage{amsmath}
\usepackage{appendix}
\usepackage{xspace}

\usepackage{epsfig}

\newtheorem{thm}{Theorem}[section]

\newtheorem{example}[thm]{Example}

\newcommand{\qed}{\hfill $\blacksquare$}

\newcommand {\lieG} {SU(2^n)\xspace}
\newcommand {\lieg} {\mathfrak{w}\xspace}
\newcommand {\controlSet} {\mathcal{V}\xspace}

\newcommand {\controlSetForExample} {\controlSet}
\newcommand {\controlVFBound} {C_{\max}\xspace}
\begin{document}
\title{Gate complexity using Dynamic Programming}
\author{Srinivas Sridharan}
\affiliation{Department of
    Engineering, Australian
    National University, Canberra, ACT 0200,
    Australia.}
\email[E-mail address to which correspondence should be sent: ]{srinivas.sridharan@anu.edu.au}
\author{Mile Gu}
\affiliation{Department of Physics, University of Queensland,
Brisbane, QLD, 4072, Australia.}
\author{Matthew R.~James}
\affiliation{Department of
    Engineering, Australian
    National University, Canberra, ACT 0200,
    Australia.}

\date\today
\begin{abstract}
The relationship between efficient quantum gate synthesis and
control theory has been a topic of interest in the quantum control
literature. Motivated by this work, we describe in the present
article how the dynamic programming technique from optimal control
may be used for the optimal synthesis of quantum circuits. We
demonstrate simulation results on an example system on $SU(2)$, to
obtain plots related to the gate complexity and sample paths for
different logic gates.
\end{abstract}
\pacs{03.67.Lx, 02.70.-c}
\maketitle

\section{Introduction}
One of the important pursuits in the field of quantum computation is
the determination of computationally efficient ways to synthesize
any desired unitary gate from fundamental quantum logic gates. Of
special interest are  the bounds on the number of one and two qubit
gates required to perform a desired unitary operation (termed the
gate complexity of the unitary). This may be considered a measure of
how efficiently an operation may be implemented using fundamental
gates.

An  approach linking efficient quantum circuit design to the problem
of finding a least path-length trajectory on a manifold was taken up
in \cite{nielsen2006qcg}; the path length was related to minimizing
a cost function to an associated control problem (with
\textit{specific} Riemannian metrics as the cost function). This
approach was later generalized in \cite{Nielsen2006} to use a more
general class of Riemannian metrics as cost functions to obtain
bounds on the complexity. In \cite{schulteherbruggen2005ocb} the
authors use Pontryagins' maximum principle
 from optimal control theory to obtain a minimum time implementation of
  quantum algorithms.~Alternative techniques using Lie group
decomposition methods to obtain the optimal sequence of gates to
synthesize a unitary were developed in
\cite{dalessandro2002ufg,schirmer2001qcu,ramakrishna2000qcd}.

 In this article we use the method of dynamic programming
from the theory of optimal control to determine the sequence of one
and two qubit gates which implement a desired unitary. We solve the
problem using one and two qubit Hamiltonians as the control vector
fields, in contrast to the approach in \cite{nielsen2006qcg} which
used the concept of \lq preferred\rq \, Hamiltonians. In addition we
demonstrate numerical results on an example problem in $SU(2)$ and
obtain controls which may be then be used to split any gate into a
product of fundamental unitaries.

The organization of this article is as follows. In Section
\ref{sec:preliminaries} we provide a review of the definitions of
gate complexity, approximate gate complexity and the control problem
associated with the bounds on these quantities. We describe, in
Section \ref{sec:dynProgGateComplexity} how this control problem can
be solved using Dynamic Programming techniques from mathematical
control theory. This is followed by a demonstration, in Section
\ref{sec:exampleProblem}, of the theory developed to an example
problem on $SU(2)$; wherein simulation results and sample optimal
trajectories are obtained.

\section{Preliminary Concepts} \label{sec:preliminaries}
In this section we recall the notion of gate complexity and its
relation to the cost  function for an associated control problem as
in \cite{Nielsen2006}.

\subsection{Gate complexity}\label{subsec:gateComplexity}
As outlined in \cite[Chapter 4]{nielsen2000qca}, in quantum
computation each desired quantum algorithm may be defined by a
sequence of unitary operators $U_1\,,\,U_2\,,\ldots$. Each of these
$U_j$ is an element of the Lie group $SU(2^n)$ and represents the
action of the algorithm on an $n$-qubit input. The gate complexity
$G(U_0)$ of a unitary $U_0$ is the minimal number of one and two
qubit gates required to synthesize $U_0$ exactly, without help from
ancilla qubits \cite{Nielsen2006}. The complexity of the algorithm
is a measure of the scaling of the amount of basic resources
required to synthesize the algorithm, with respect to input size.

In practice however, computations need not be exact. To perform a
desired computation $U_0$, it may suffice to synthesize a unitary
 $\hat{U}_0$ with accuracy $\epsilon$, i.e $\|U_0-\hat{U}_0\| \leq \epsilon$. Here
  $\|\cdot\|$ denotes the standard matrix norm in a particular representation of the group.
This notion gives rise to the definition of the approximate gate
complexity $G(U_0,\epsilon)$ as
  the minimal number of one and two qubit gates required to
  synthesize $U_0$ up to accuracy $\epsilon$.

\subsection{Control problem}\label{subsec:ControlProblem}
We outline below a control problem on the Lie group $SU(2^n)$ such
that the cost function associated with it provides upper and lower
bounds on the gate complexity problem. The system evolution for the
control problem occurs on $\lieG$ with associated Lie algebra
$\lieg~=~\mathfrak{su}(2^n)$. Note that in this article the definition
of $\mathfrak{su}(2^n)$ is taken to be the collection of traceless
Hermitian matrices (which differs from the mathematicians'
convention by a factor of $i$). The system equation contains a set
of right invariant vector fields $H_1,\,H_2\,\ldots\,H_m$ which
correspond to a set of one and two qubit Hamiltonians. The Lie
algebra generated by the set $\{ \, H_1,\,H_2\,\ldots\,H_m\, \}$ is
assumed to be $\lieg$. This assumption along with the fact that
$\lieG$ is compact imply that the system is controllable
\cite{JURDJEVIC}(and hence the minimum time to move between any two
points on $\lieG$ is finite).
  The system dynamics
for the gate design problem is described as follows:
\begin{align}
\frac{dU}{dt} = -i\,\{\sum_{k=1}^m v_k(t) H_k\} &U,\qquad
 U\,\in\,\lieG \label{eq:System}
\end{align}
with the initial condition $U(0)=U_0$ and with a  bound  $\controlVFBound$ on the norm of the available control vector
fields $H_j$ using any suitable norm on their matrix representation.
The control $v$ is an element of the class of piecewise continuous
functions with their range belonging to a compact subset
$\mathbf{V}$ of the real $m$-dimensional Euclidian space
($\mathbb{R}^m$). We denote this class of functions by
$\controlSet$. Hence $\controlSet:=\{v(.)\,|\,\forall\,i\,\,
v_i\,:\,[0,\infty\,)\,\rightarrow\,\mathbb{R}\,,\,v_i\,\mathrm{\,is\,piecewise\,continuous}\}$.

Given a control signal  $v$ and an initial unitary $U_0$ the
solution to Eq~\eqref{eq:System} at time $t$ is denoted by
$U(t;v,U_0)$. In addition, by a simple time reversal argument it can be seen that
the problem of obtaining a desired unitary gate $U_0$ starting from
the identity can be reframed as a problem of reaching the identity
element starting at $U_0$.

Note the difference between the system described herein  and that
outlined in the control problem in \cite{Nielsen2006}. In the
present case, the control Hamiltonians used are the ones which
generate the one and two qubit unitaries; therefore the concept of
\lq preferred\rq\,  vs \lq allowed\rq\,  Hamiltonians is not
utilized here. The controls used  may in fact be any
bracket-generating subset of  the vector fields which generate one
and two qubit unitaries. 

We now define some terms to be used in this article: For $U_0\,\in
\lieG$ and $v \in \controlSet$
\begin{description}
  \item[Time to reach the Identity using control $v$]
  \begin{align}
t_{U_0}(v) = \inf \{ t > 0 \ : \ U(0) = U_0, \nonumber\\U(t)=I, \
\mathrm{dynamics\,in\,}  \mathrm{(\ref{eq:System})} \} .
\label{tUv-def}
\end{align}
The infimum in Eq \eqref{tUv-def} is infinite if the terminal
constraint $U(t)=I$ is not attained.
  \item[Cost Function]
\[
  J(U_0,v) := \int\limits_0^{t_{U_0}(v) } {\ell(v(s))} ds,
\]
given the dynamics in Eq \eqref{eq:System} with control
$v\,\in\,\controlSet$ where $\ell\,:\,\mathbf{V}
\rightarrow\,\mathbb{R}$ is continuous and has a finite positive
maximum and minimum i.e $L_{min}~\leq~\ell(v)~\leq~L_{\max}$. Note that
as long as $\ell$ has a minimum greater than zero, the bounds on
it can be recast in the form $1\leq\,L_{min}\leq\ell(v)$.

\item[Optimal Cost Function]
\begin{align}
  C(U_0) &= \mathop {\inf
  }\limits_{v\in\controlSet}{J(U_0,v)}\label{eq:CostFunctionGeneral}
  \end{align}
This optimal cost function will be used to provide bounds on the
gate complexity.
\end{description}

Hence the control problem is to find the values of $v$ in order to
optimize the cost function. The boundedness of the time taken to
achieve the desired objective (due to controllability), together
with
 the boundedness of $\ell$, implies that the cost function is bounded. In addition, the
control problem may also be generalized to systems evolving on any
compact connected Lie Group $\mathcal{M}$ with the cost $\ell(x,v)$
being dependent on both $x\,\in\,\mathcal{M}$ and
$v\,\in\,\controlSet$.

\subsection{Bounds relating gate complexity and control}
\label{sec:GateComplexityBounds} We now recall results on the
relation between the cost of the associated control problem and both
the  upper bound on the approximate gate complexity and the lower
bound on the gate complexity \cite{nielsen2006qcg,Nielsen2006}.

We define
\begin{eqnarray}
T_{\max}:=\mathop {\max }\limits_{U\in \mathcal{U}_{1/2}} \{ \mathop
{\inf }\limits_{v\in\controlSet} [t_U(v)]\},
\end{eqnarray}
where $\mathcal{U}_{1/2}$ is the set of one and two qubit unitary
gates. Hence the total time to construct $U_0$ from $I$ or vice-versa is
at-most $G(U_0)\times T_{\max}$. Therefore for any element $U_0$ of
$\lieG$ we have \cite{nielsen2006qcg} 
\begin{align}
C(U_0)\leq L_{\max}\,G(U_0)\,T_{\max}.\label{eq:lowerBound}
\end{align}

From \cite{Nielsen2006} we have that a given unitary $U_0$ in
$SU(2^n)$ can be approximated to $O(\epsilon)$ using $O(C(U_0)^3
n^6/{\epsilon}^2)$ one and two qubit unitary gates. Hence the upper
bound on the approximate gate complexity satisfies
\begin{align}
G(U_0,{\epsilon})\leq O(\frac{n^6\,C(U_0)^3}{{\epsilon}^2}).
\label{eq:upperBound}
\end{align}
 This motivates the solution to certain related optimal control problems in order to obtain bounds on the
complexity of related quantum algorithms. In addition,
the solutions to such optimal control problems help determine the sequence of
one and two qubit gates used to generate the desired unitary as
described in the following section.

\section{Dynamic Programming for gate complexity}
\label{sec:dynProgGateComplexity}
In this section we introduce the tools of dynamic programming which
have had widespread application in control theory. We then apply
this theory to solve the control problem associated with
determination of the bounds on gate complexity and explain how to
use the solution to the control problem to obtain the sequence of
gates required to reach any given unitary.

\subsection{Introduction to Dynamic Programming}
The dynamic programming principle states that: \lq An optimal policy
has the property that, whatever the initial state and optimal first
decision may be, the remaining decisions constitute an optimal
policy with regard to the state resulting from the first decision\rq
\cite[p.83]{bellman2003dp}. The use of this principle involves
recursively solving a problem by breaking it down into several
sub-problems followed by determining the optimal strategy for each
of those sub-problems. For example, consider the following problem:

\begin{example}\label{example:DPP}
\end{example}
Let the state of a system be described by a  vector in some vector
space $\mathcal{M}$. Moving from one point to another is done by
exerting a control chosen from a compact control set $\mathbf{V}$
(we assume full controllability of the system using controls from
this set).
Any control $v$ exerted starting at a point $U$ in this space leads
to a point denoted as $P(v,U)$, while incurring a cost. This cost of
exerting the control may depend on the point at which the control
starts being applied as well as on the control signal itself. The
objective is to reach from a point $A$ to a point $B$ in that vector
space while incurring as low a cost as possible. Let the cost of
reaching $B$  from any point $U$ using control $v$ be denoted by
$J(U,v)$. The cost at any point is denoted by $C(U)\,,
\,U\,\in\,\mathcal{M}$.
%
The dynamic programming principle implies:
\begin{align}
C(U)\leq \mathop {\inf}\limits_{v\,\in\,\controlSet} \{ C(P(v,U)) +
J(U,v)\}.\label{eq:lowerLimitonDPEexample}
\end{align}
Also  from the definition of $C(U)$ we have that for any
$\epsilon\,>\,0$, there exists a control $v^*$ s.t
\begin{align}
C(U)+\epsilon \geq C(P(v^*,U)) + J(U,v^*).
\end{align}
Hence,\begin{align}
C(U)+\epsilon \geq \mathop {\inf}\limits_{v\,\in\,\controlSet} \{
C(P(v,U)) + J(U,v)\}.
\end{align}
In the limit $\epsilon\rightarrow 0$ we
have that:
\begin{eqnarray}
C(U) \geq \mathop {\inf}\limits_{v\,\in\,\controlSet} \{ C(P(v,U)) +
J(U,v)\}. \label{eq:upperLimitonDPEexample}
\end{eqnarray}
Equations \eqref{eq:lowerLimitonDPEexample} and
\eqref{eq:upperLimitonDPEexample}  imply the dynamic programming
equation
\begin{eqnarray}
C(U) = \mathop {\inf}\limits_{v\,\in\,\controlSet} \{ C(P(v,U)) +
J(U,v)\}.  \label{eq:example_DPE}
\end{eqnarray}
\qed

Solving a control problem using this principle involves setting up
and solving a recursion equation as above. There are several
references which provide a detailed and rigorous introduction to
this theory viz \cite{bertsekas1995dpa,bryson1975aoc,kirk2004oct}.
We now apply this theory to obtain bounds on the gate complexity.

\subsection{Use in the gate complexity
problem}\label{subsec:UseofControlinGatecomplexity}

By the  procedure described in Example (\ref{example:DPP}) in the
previous sub-section, we have that the dynamic programming equation
for the optimal cost function in Eq \eqref{eq:CostFunctionGeneral}
is
\begin{align}
C(U_0) =&\mathop {\inf }\limits_{v\,\in\,\controlSet} \left\{
\int\limits_0^{t \wedge t_{U_0} (v)}{\ell(v(s))ds} + \right.
\nonumber
\\&\left. \chi_{t<t_{U_0} (v) } C(U(t \wedge t_{U_0} (v);v,U_0 ))
\right. \Bigg{\}}, \label{eq:DPEGeneral}
\end{align}
for all initial points $U_0\in \lieG$. Here $a \wedge b : = \min\{a,b\}$ and  $\chi_{\Omega}$ is the
indicator function of the set $\Omega$ taking on the value of $1$
inside the set and zero outside it. Thus, $\chi_{t<t_{U_0}}$ is $1$
if $t<t_{U_0}$.

Now we describe a formal derivation to obtain the differential
version of the dynamic programming equation(DPE). For sufficiently
small $t$ we have from Eq \eqref{eq:DPEGeneral} that:
\begin{align}
C(U_0) =\mathop {\inf }\limits_{v\,\in\,\controlSet} \left\{
\int\limits_0^{t}{\ell(v(s))ds} + C(U(t;v,U_0 )) \right\}
\label{eq:dpeForHJB_1}
\end{align}

Now transposing $C(U_0)$ to the right hand side, dividing by $t$ and
taking the limit as $t\rightarrow 0$ we obtain:
\begin{align}
0 =&\mathop {\inf }\limits_{v\,\in\,\mathbf{V}} \left\{ \ell(v) +
DC(U_0)(\dot{U})\right\}\\
0 =&\mathop {\inf }\limits_{v\,\in\,\mathbf{V}} \left\{ \ell(v) +
DC(U_0)\left[-i\,\{\sum_{k=1}^m v_k(t) H_k\} x\right]\right\}\\
0 =&\mathop {\sup }\limits_{v\,\in\,\mathbf{V}} \left\{-\ell(v) -
DC(U_0)\left[-i\,\{\sum_{k=1}^m v_k(t) H_k\} x\right]\right\}.
\end{align}
In the equation above $DC(U_0)$ denotes the derivative of the
function $C$ at a point $U_0$ on the Lie group $\lieG$. Hence the function $C$ (Eq
\eqref{eq:DPEGeneral}) satisfies
\begin{align}
\sup \limits_{v \in\mathbf{V}}\left\{-\ell(v)-\right.
&\left.DC(U)\left[-i\,\{\sum_{k=1}^m v_k(t) H_k\}
U\right]\right\}=0, \label{eq:HJBgeneralresult}
\end{align}
termed the Hamilton-Jacobi-Bellman (HJB) equation. In principle this
solution $C$ can be used to obtain bounds on the gate complexity as
indicated in Eqns \eqref{eq:lowerBound} and \eqref{eq:upperBound}.
Assuming regularity conditions,  the optimal control policy is
generated by the synthesis equations given below \cite[Section
1.5]{Bardi}. $v^*$ is optimal for an initial state $U_0$ if and only
if $v^*(t) \in R(U(t))$ for all $t>0$ (almost everywhere), where
\begin{align}
 R(U)&:=  \underset{v\,\in\,
\mathbf{V}}{\operatorname{argmax}}\,\left\{ -DC(U)\left[-i\,
\{\sum_{k=1}^m v_k(t) H_k\}
U\right]-\ell(v)\right\}.\label{eq:GeneralcontrolSynthesis}
\end{align}
In these expressions we define $U(t):=U(t;v,U_0)$ to be the solution to the differential equation
Eq \eqref{eq:System} at time $t$ with control history $v$.

\subsection{Obtaining one and two qubit gate implementations}
\label{sec:obtainingGateImplementations}
We numerically synthesize
the optimal controls using techniques such as in \cite[Chapter
3]{kushner2001nms} where the solutions to the discretized version
tend towards the solution of the original continuous description.

The solution to the HJB gives the control sequence to be applied to
reach the identity element. This is the crucial step in the control
synthesis.
From the Baker-Campbell-Hausdorff formula \cite{hall2003lgl} we know
that
\begin{align}
\exp\{i\,(A+B)\,\Delta t\}&= \exp\{i\,A\,\Delta
t\}\times\exp\{i\,B\,\Delta t\}+O({\Delta t}^2),
\label{eq:smjpra2008:BCH}
\end{align}
where $A$ and $B$ are any Hermitian operators. This implies that in
a one qubit system, a unitary generated by any element of the Lie
algebra of a one qubit system can be approximated as closely as
desired by the product of unitaries generated by flowing along the
available one and two qubit control Hamiltonians (given the bracket
generating assumption mentioned previously). We now recall two
statements regarding the universality of gates and gate synthesis
using a product of two level unitaries below.
\begin{enumerate}
  \item A single qubit and a C-not gate are universal i.e produce
  any two level unitary  \cite[Section 4.5.2]{nielsen2000qca}.
  \item An arbitrary unitary matrix on a $d$-dimensional Hilbert
  space can be written exactly as a product of two level unitaries \cite[Section 4.5.1]{nielsen2000qca}.
\end{enumerate}

These statements together with Eq \eqref{eq:smjpra2008:BCH} indicate
that once we have a control vector from the solution of the HJB
equation, it is possible to synthesize a one and two qubit gate
sequence to approximate a desired unitary to as good an accuracy as
required.

\section{Example problem on $SU(2)$}\label{sec:exampleProblem}
We now use the theory introduced to consider an example on the
special unitary group. We wish to construct any element of $SU(2)$
using the available Hamiltonians $I_x$ and $I_z$. The system
dynamics on $SU(2)$ is given by:
\begin{align}
\frac{dU}{dt} &= -i\,(v_1\,I_x+v_2\,I_z)U,\qquad  U\,\in\,SU(2)
\label{eq:exampleSystemSu2}
\end{align}
where $v \,\in\,\controlSetForExample$ and
\begin{align}
I_x&= \frac{1}{2}\,\left(
  \begin{array}{cc}
    0 & 1 \\
    1 & 0 \\
  \end{array}
\right)\quad I_z = \frac{1}{2} \,\left(
  \begin{array}{cc}
    1 & 0 \\
    0 & -1 \\
  \end{array}
\right).
\end{align}
In this case $\ell(.):=1$.

  This cost function in effect measures the distance along the
  manifold to generate the desired unitary. Due to the fact that cost function does not depend
  on the magnitude of the control signal applied, the problem essentially involves choosing
  the direction to flow along (with maximum magnitude), at each point on the manifold in order to reach the destination in the
  smallest possible time. Hence the direction (and thus the path)
is chosen in order to minimize the the \lq distance\rq \,along the manifold.
  Thus this minimum time control problem is related to the original gate complexity
  problem. The minimum time problem in quantum mechanics has also
  attracted interest in other articles
  \cite{N.2001,boscain2005tos,schulteherbruggen2005ocb,carlini2006toq}.

The HJB for the cost function is:
\begin{align}
\sup \limits_{v \in
\mathbf{V}}\left\{-1-DC(U)\left[-i\,(v_1\,I_x+v_2\,I_z)U
\right]\right\} =0\label{eq:HJBExamplesu2}
\end{align}
where $U\,\in\,SU(2)$. The optimal control is chosen according to
the Eq \eqref{eq:GeneralcontrolSynthesis}.

Now, in order to obtain the numerical solution to this problem we
proceed as follows. Instead of using the value function $C$
directly, it is advantageous to use the monotone transformation
(Kruskov transform)
\begin{equation}
S(U) = 1-e^{-C(U)}  \,, \,\quad U \in SU(2) \label{def:S:su2example}
\end{equation}
which leads to the following HJB equation (using \cite[Proposition
2.5]{Bardi}):
\begin{eqnarray}
S(U)+H(U, DS(U)) &=& 0
\nonumber\\
1 \geq S(U) & > & 0  ,\,\,\, U\,\in\,SU(2)\setminus I
\label{eq:HJBdiscounted:su2example}\\
S(I) &=& 0 \,\,\mathrm{(\,boundary\,condition\,)}\nonumber
\end{eqnarray}
with the Hamiltonian term $H$ being the same as in
Eq~\eqref{eq:HJBExamplesu2}. The function $S$ can be interpreted as
a discounted minimum time function for the system in
Eq~\eqref{eq:exampleSystemSu2}. Therefore from the Dynamic
Programming principle $S$ would satisfy
\begin{align}
S(U_0) &=\mathop {\inf }\limits_{v\,\in\, \controlSetForExample }
\left\{
\int\limits_0^{t \wedge t_{U_0}(v)} {e^{ - s} ds} + \right.\nonumber\\
&\left. e^{ - (t \wedge t_{U_0}(v))} S(U(t \wedge t_{U_0}(v);v,U_0))
\right.\mathrm{}\Bigg{\}} \label{eq:DppDiscountedExamplesu2}.
\end{align}

This normalization (discounting) is useful for  better numerical
convergence and is also used in the uniqueness proofs of the
solutions to the dynamic programming equations.

To obtain a numerical solution to the dynamic programming problem,
we parameterize points in $SU(2)$ using a mapping of the
form $\exp(a\,I_x+b\,I_y+c\,I_z)$ from the Euclidian space. Note that
the parametrization is not unique, since multiple points in
the three dimensional Euclidian space $\mathbb{R}^3$ map to the same
point in $SU(2)$.

Note that since HJB equation
in this case is linear in $v$, the optimal $v$ lies on the
boundary of the compact set $\mathbf{V}$ at (almost) every time
instant. This simplification reflects the results from
\cite{nielsen2006qcg} where determining the geodesic (which are
paths of constant magnitude of the velocity) involves choosing an
optimal direction along which to flow.

\begin{figure}[hpt]
\centering
\includegraphics[scale=.4]{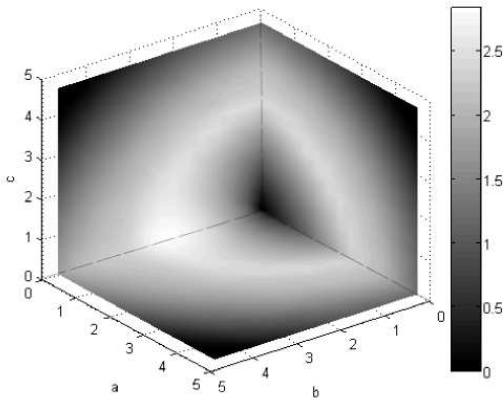}
\caption{Un-normalized Optimal cost function with a control of norm
1 }\label{fig:timeComparisonSU2ckt}
\end{figure}

The method of discretization of
Eqns~\eqref{eq:HJBdiscounted:su2example}, which are used to obtain
the simulation results, will be described later in this article.
Figure \ref{fig:timeComparisonSU2ckt} indicates the slices along a
quadrent of the co-ordinate axes of the actual minimum time function
$C$ (which corresponds via Eq~\eqref{def:S:su2example} to the
normalized minimum function $S$, obtained by solving the HJB
Eq~\eqref{eq:HJBdiscounted:su2example}).

The figure is presented as a gray-scale image in a three dimensional
grid. The axes correspond to the three parameters used for the
representation of $SU(2)$ as described above. A lighter shading
indicates a larger value of the minimum time function at a point,
while a darker shading implies a smaller time to reach the identity
element when starting from that point.

\begin{figure}[hpt]
\centering
\includegraphics[scale=.6]{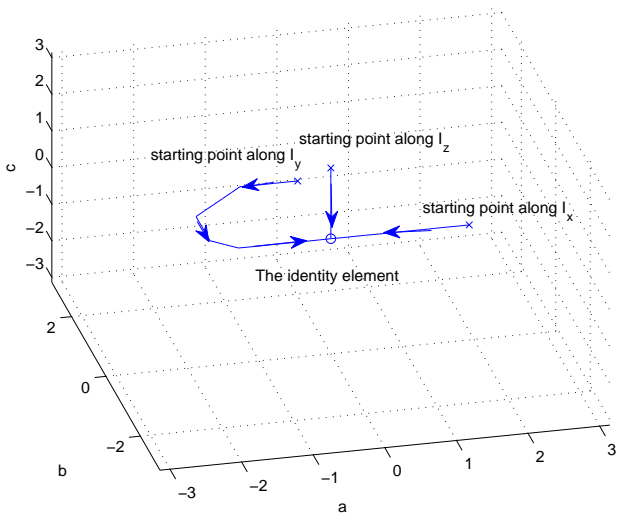}
\caption{Path to the identity element starting at different points
on each of the 3 axes.
 }\label{fig:mintimePathExampleSU2}
\end{figure}

Time optimal trajectories for this example are indicated in Figure
\ref{fig:mintimePathExampleSU2}.  Note that the non-uniqueness of
the representation leads to having to carefully interpret the paths
when they are shown in flat space. Observe that since there
is no direct vector field to control along $I_y$,
the path to the identity starting from a point along $I_y$ is not a
straight line unlike in the case of the other two axes ($I_x$ and
$I_z$).

The discretization of this system for obtaining numerical solutions
to the HJB equation (\ref{eq:HJBdiscounted:su2example}) is carried
out using the finite difference procedure in \cite[Section
6.5]{kushner2001nms}. In three dimensional Euclidian space with a
grid spacing $h$ of the space and basis vectors $e_i$, the value
iteration equation
 (which is the iteration of the cost function, say $S^{h}$~) is given by:
\[
{S^{h}}(x) = \mathop {\inf }\limits_U \left\{ {\frac{h}{{h + \left\|
f \right\|_1 }} + \frac{{\sum\limits_{i = 1}^3 {{S^{h}}_ \pm ^i
(x)\,f_ \pm ^i (x,U)} }}{{h + \left\| f \right\|_1 }}} \right\},
\]
where
\begin{align}
{ S^{h}}_ \pm ^i  &:= {S^{h}}(x \pm h\,e_i ) \nonumber\\
 f_ + ^i (x,U) &:= \max \left\{ {f^i(x,U),0} \right\} \nonumber \\
 f_ - ^i (x,U) &:=  - \min \left\{ {f^i(x,U),0} \right\}
 \nonumber\\
\left\| f(x) \right\|_1&:=\sum\limits_{i = 1}^n {\left\| f^i(x)
\right\|} \label{equation:numericalSolnDppAdj}.
 \end{align}
Here  $f^i$ are the $i$ th components of the vector valued function
$f$.

Note that the optimal control for this system is a specific case of
Eq~\eqref{eq:GeneralcontrolSynthesis}, where the possible values of
the spatial co-ordinates are the locations of the grid points which
in turn depend on the mesh generated for the discretization. These
discretized equations and the controls resulting therefrom are used
to obtain the simulation results indicated in the figures in this
article. Once the
 controls are determined, we can generate the one and two qubit
 unitaries which efficiently approximate this control trajectory as explained in Section
 \ref{sec:obtainingGateImplementations}.

\section{Discussion and conclusions}
In this article we have described the use of the Dynamic programming
method to solve the efficient gate synthesis problem and have
demonstrated a proof of principle of this technique by obtaining a
complete solution to an example problem of a single qubit.

A comparison between the method introduced and algebraic
decomposition based approaches (such as applications of the methods
in \cite{N.2001}), is shown in \cite{sjcdc2008Underreview}; wherein it is
demonstrated that the
results obtained by a decomposition based method agree well, to
within the  error bounds of the discretization, with those resulting
from the dynamic programming based control method.

The methods in the present article are sufficiently general to be
able to be used with various cost functions such as the ones in
\cite{schulteherbruggen2005ocb} as well those used in geometric
approaches to the problem as in \cite{Nielsen2006}.

The simulations in this work are based on theoretical results which
are quite involved. A rigorous and complete development of the
proofs of the foundations of this article will be deferred to a
future publication. The numerical procedures outlined herein
generalize to higher dimensional cases with the crucial limiting
factor being the time taken and storage requirements for these
computations (which increases dramatically with the dimension of the
system). The treatment of problems of direct interest to gate
complexity will require an analysis of unitaries on three or more
qubits. Owing to the curse of dimensionality, further work is
required to develop computational methods of greater efficiency in
order to use the Dynamic Programming technique to investigate these
problems of practical interest.

\begin{acknowledgments}
S.~Sridharan and M.R.~James wish to acknowledge the support for this work by the
    Australian Research Council.
\end{acknowledgments}

\end{document}